# Extreme Linewidth Narrowing in Diamond Raman Lasers Enables the Generation of 35 W at 589 nm with Hz-Scale Intrinsic Linewidth

OSAMA TERRA[1,*], ADAM SHARP[1], AIDAN CONNAUGHTON[1], MARK FERRIER[1], JIPENG LIN[2], TIAGO A. ORTEGA[2], DAVID J. SPENCE[1], RICHARD P. MILDREN[1]

[1] *School of Mathematical and Physical Sciences, Macquarie University, NSW 2109, Australia*
[2] *EOS Space Systems Pty Ltd, 18 Wormald Street, Symonston, ACT 2609 Australia*
*[*osama.terra@mq.edu.au](mailto:osama.terra@mq.edu.au)*

**Abstract:** High-power lasers with ultranarrow linewidth and high beam quality in the visible spectrum are essential for emerging quantum and space technologies. Here we report significant advances in diamond Raman lasers, generating diffraction-limited yellow light at 589 nm with output power up to 35 W and enhanced single-frequency stability. The optical-to-optical efficiency from the pump reaches 47.7%, representing a record efficiency for this class of devices. More importantly, the extreme linewidth-narrowing of Raman lasing in diamond enables a reduction in frequency noise exceeding six orders of magnitude, resulting in a measurement-limited intrinsic linewidth of 6 Hz at the maximum power. The laser is further stabilized to the sodium $D_{2a}$ saturation-absorption transition, making it well-suited for sodium-based space and quantum experiments. These results represent a major step toward continuous-wave, high-power, single-frequency laser sources across the visible spectrum that combine ultranarrow linewidth, high efficiency, and near-diffraction-limited beam quality for advanced quantum and space applications.

## 1. Introduction

High-power visible lasers with ultranarrow linewidths and near-diffraction-limited beam quality are essential for emerging quantum technologies and space-based applications. Many advanced quantum platforms, including optical lattice clocks, atom interferometers, and Rydberg atom arrays, are restricted in their choice of atomic species due to the limited availability of laser sources that precisely match key atomic transitions or optical lattice wavelengths [1]. Access to higher optical power enables larger atomic ensembles and deeper trapping potentials, thereby improving the sensitivity of quantum sensors and increasing the number of addressable atoms or qubits in quantum computing architectures [2]. Ultranarrow integrated and intrinsic linewidths are critical for maintaining long coherence times and achieving high-fidelity control and gate operations in these systems [3] [4].

For space applications, high optical power and near-diffraction-limited beam quality, together with stable single-frequency operation, are particularly demanding requirements. Lasers at 589 nm are required to resonantly excite sodium atoms in the mesospheric layer (~90 km altitude). The excited atoms subsequently re-emit photons via resonant fluorescence to form an artificial laser guide star used for adaptive optics. The guide star provides a reference wavefront that allows real-time correction of atmospheric turbulence, which significantly improves the angular resolution of ground-based telescopes [5]. This capability also supports space situational awareness, including improved imaging and tracking of space debris [6], and enhanced free-space optical communication with satellites [7]. Other space-related applications at this wavelength include mesospheric sodium magnetometry and sodium-layer remote sensing, which require narrow-linewidth lasers to perform high-resolution spectroscopic measurements, resolving small frequency shifts caused by geomagnetic Zeeman splitting or Doppler shifts from mesospheric winds [8][9].

Four notable laser technologies have been developed to reach this challenging wavelength with such power, spectral, and spatial qualities. Early systems utilized dye lasers and sum-frequency generation, but dye lasers require high maintenance and precise synchronization of 1319 nm and 1064 nm sources, resulting in large, complex, and less reliable setups [10]. Fiber Raman amplification of a 1178 nm seed laser, followed by free-space frequency doubling to 589 nm, has emerged as an alternative [11]. This approach has produced 20 W commercial systems and recently achieved 74 W using cascaded Raman amplification with 248 W pump power, with an efficiency of 29.8% and a linewidth of 2.3 MHz at 25 W [12] [13]. Recent advances in VECSELs and MECSELs have produced up to 15 W at 589 nm with a few-MHz linewidths [14], [15]. Therefore, these systems are primarily suited to guide-star applications.

Diamond Raman lasers (DRLs) are an emerging technology with power-scaling potential, enabled by diamond's exceptional thermal conductivity (~2000 W/m·K). Unlike population-inversion gain, Raman gain depends on the pump intensity rather than a stored inversion, making it less susceptible to spatial hole burning. This allows single-mode operation in standing wave resonators without intra-cavity mode selectors [16]. DRLs also offer a simpler architecture than fiber Raman amplifiers by enabling intracavity second-harmonic generation (SHG), allowing visible wavelengths to be generated within the same laser cavity [17][18][19]. More importantly, numerical modeling by our group predicts that Raman lasing can achieve linewidth narrowing of up to eight orders of magnitude, which, if experimentally realized, would surpass linewidth reduction techniques such as Brillouin lasers and self-injection locking by several orders of magnitude [20], [21][22].

Despite these advantages, several challenges remain in diamond Raman lasers that limit the full realization of their potential. Broadband antireflection (AR) coatings on diamonds are particularly challenging due to their high refractive index, chemical inertness, and damage risk from the tightly focused beams. In practice, only a narrowband coating at the Stokes wavelength is feasible. Folded cavities, such as the V- and Z-shaped, are used to separate the frequency-doubled light from the arm of the diamond crystal [17]. These cavities use a spherical turning mirror to introduce another focus in the SHG crystal to allow an efficient SHG, but the tilted spherical mirror causes astigmatism in the cavity. This leads to instabilities from the unavoidable distributed SBS-assisted higher-order spatial modes throughout the cavity's free-spectral range [23].

Here, we report a single-focus L-shape cavity design for intracavity-SHG diamond Raman lasers allowing a record optical-to-optical efficiency of 47.7% and continuous-wave light of 35 W at 589 nm with near-diffraction-limited beam quality ($M^2 = 1.07$). Moreover, leveraging the linewidth narrowing of Raman lasers, an unprecedented six orders of magnitude frequency noise reduction has been achieved, delivering a 6 Hz intrinsic linewidth at the maximum power. The linewidth narrowing enables the use of a seed with an intentionally broadened Gaussian linewidth of ~1.5 GHz to suppress SBS in the high-power fiber amplifier, resulting in a laser with a Gaussian linewidth of only 126 kHz at 35 W of yellow light, limited by the technical noise in the laser cavity. This enables, for the first time, the stabilization of the diamond Raman lasers to the $D_2a$ sodium hyperfine transition using a dual-piezo symmetric frequency control scheme.

**Results**

## 2. Laser cavity design

Fig. 1(a) shows the single-focus L-shape cavity geometry, which was chosen to allow extraction of the SHG output in a single beam without passage through the diamond. Placing the diamond in a different arm from the path of the yellow light generated by the LBO significantly improves the beam quality by eliminating the interference caused by the diamond faces at this wavelength (Fig. 1(b)). Two strategies enable optimal output coupling without an additional focus in the LBO crystal. The L-shape geometry allows placing the LBO crystal close to the focus in the diamond to enhance SHG efficiency. In addition, the double-pass SHG, achieved via yellow reflection from the input coupler, provides up to a fourfold enhancement, in tandem with our recent dephasing study [24]. The flat mirror avoids the astigmatism caused by the tilted spherical mirror, which limits the spread of the SBS-assisted higher-order spatial modes over the cavity FSR and allows SBS-free and stable regions (see methods section for more details).

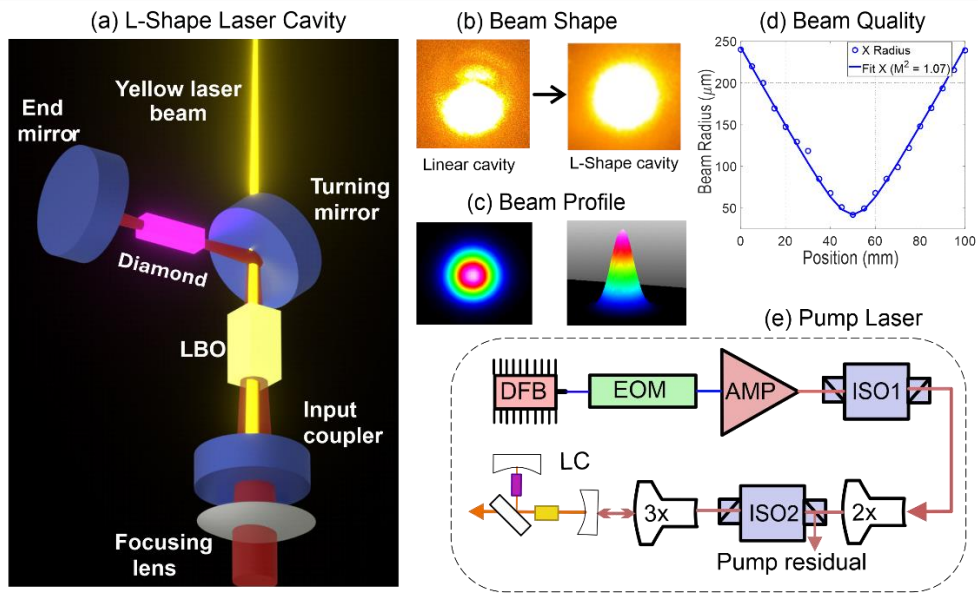

Fig. 1. (a) Single-focus L-shaped laser cavity. Pump light passes the frequency-doubling crystal (LBO), and is focused into the diamond in the orthogonal arm. (b) Beam shape comparison between linear to L-shaped cavities, showing elimination of interference fringes by the L-shape cavity, (c) Beam profiles: L-shaped cavity output with ellipticity < 3%, (d) Beam quality measurement using a scanning-slit profiler, showing an $M^2$ value of 1.07. (e) Pump laser layout: a DFB seed laser, electro-optic phase modulator (EOM), Yb-doped fiber amplifier (AMP), two first isolators (ISO1, ISO2), two beam expanders (2, 3X), and the laser cavity (LC) (more details in the methods section).

The folded design is configured to provide very low loss on the reflected s-polarization, which assists in increasing power and efficiency. Fig. 1 (c) shows the Gaussian circular beam profile from the L-shape cavity with an ellipticity of less than 3%. Fig. 1 (d) shows the beam quality measurement measured with a scanning slit beam profiler, which indicates a near diffraction-limited beam quality with an $M^2$ = 1.07. Figure 1(e) depicts the pump setup, featuring a seed DFB laser whose linewidth is broadened via an EOM to suppress SBS in the fiber amplifier (more details in the methods section). Mode matching between the pump and cavity modes is ensured by selecting a suitable pump focusing lens to enhance pump depletion into the Stokes field. The linewidth of the pump is broadened to mitigate the SBS in the fiber amplifier.

## 3. Power efficiency

We achieved a continuous-wave 589 nm output with single-beam power up to 35 W, a conversion efficiency of 47.7%, and a lasing threshold of 6 W. Our single-output conversion efficiency represents a record efficiency at this wavelength compared with the state-of-the-art fiber cascaded Raman fiber amplifiers with an efficiency of 29.8% [13]. Even when compared with our own previous work on a linear cavity, in which the conversion efficiency was 17.5%×2 for dual output beams [19]. This high efficiency is attributed to the optimized output coupling, which is enabled by the double-pass SHG, providing up to a fourfold SHG enhancement, in tandem with our recent dephasing study [24]. It is further attributed to the L-shape geometry, which allows an interference-free beam while allowing controllable spacing between the LBO and the diamond and the low-loss mirrors. As shown in Fig. 2(a), the output power and the residual pump power curves are well produced by the double-pass laser model of ref. [24]. The discrepancy between simulation and experiment at higher powers is attributed to the degradation of the pump beam quality as it propagates through the diamond, with preferential on-axis depletion, an effect not captured by the plane wave models [25].

To investigate phase mismatch between the fundamental and SHG in the double-pass configuration and help optimize the output coupling, the LBO temperature was swept while monitoring the SHG output power[24]. As shown in Fig. 2(b), the two distinct SHG peaks indicate an approximately 180° phase mismatch relative to the maximum value. This mismatch arises from air dispersion, the LBO and mirror coatings, and the Gouy phase shift. Although this curve demonstrates that the nonlinear outcoupling was only 2.1 times the single-pass value rather than the theoretical maximum of 4 times, it provided near-optimal output coupling, which was the reason for enhancing the efficiency. The optimal coupling is evidenced by the relative heights of the secondary peaks and is consistent with the flattening of the residual power curve in Fig. 2(a) [18] [24].

At high power levels, efficiency is improved by ensuring good thermal heatsinking of the diamond and mitigation of stress-induced birefringence by the mount. The latter is minimized by ensuring uniform clamp pressure distribution rather than localized edge contact. Fig. 2(d) compares birefringence in a properly mounted (upper) versus improperly mounted (lower) diamond, measured using the Metropol technique [26]. For thermal management, two 2 × 2 mm heat spreader diamonds were placed in contact with the 2 × 4 mm lasing diamond to increase the surface area in contact with the copper mount. Fig. 2 (d) shows a simulation of the heat map of the diamond surrounded by copper in the case of the heat spreading configuration in comparison with a 2 × 2 mm diamond. The temperature difference with a copper 4 mm away from the diamond reaches 2 °C without heat spreading, and 1 °C with the heat-spreading scheme at the same input power and thermal contact with the copper mount. The Relative Intensity Noise (RIN) is measured at 35 W of power to reach −142 dBc/Hz between the $10^7$ and $10^8$ Hz, as shown in Fig. 2(e).

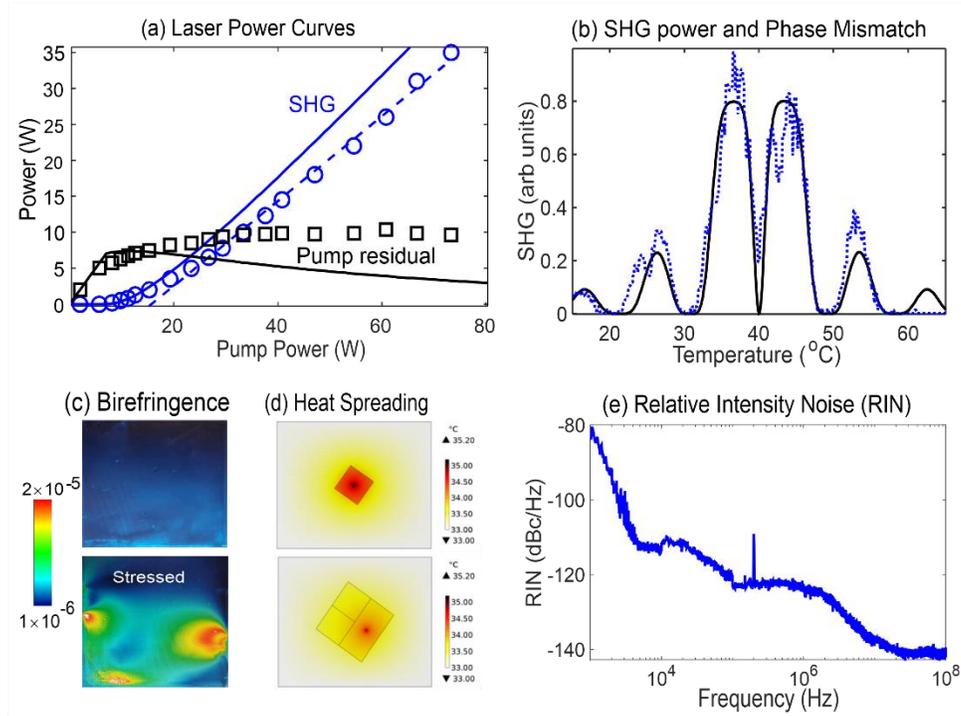

Fig. 2. (a) Output power of the diamond Raman laser, reaching 35 W with a slope efficiency of 59%, blue circles are the measurement data, blue dashed line is used for calculating the slope efficiency, blue solid line is a simulation of the model by Sharp et al [24], black squares indicate the measured residual power. (b) Temperature dependence of SHG output power, which is due to nonlinear output coupling and the phase mismatch between the Stokes and SHG fields for the second pass through the LBO crystal. (c) Birefringence measurements: properly mounted diamond with very low birefringence (~$1\times10^{-6}$); and significant birefringence increases due to improper mounting, (d) simulation of the comparison between the heat map of 2×2 lasing diamond and two 2×2 heat spreading diamonds & 2×4 lasing diamond surrounded by copper, (e) Relative intensity noise spectrum, reaching –142 dBc/Hz between $10^7$ to $10^8$.

The low relative intensity noise (RIN) is attributed to the double-pass pumping instead of employing pump resonance in the laser cavity. Although resonating the pump reduces the lasing threshold, it converts cavity frequency fluctuations into amplitude noise.

## 4. Spectral characteristics

One of the key advantages of Raman lasers is their extraordinary suppression of pump-frequency noise. The intrinsic linewidth of a laser represents its linewidth in the absence of technical noise that broadens the spectrum into a Gaussian profile. It is determined from the frequency-noise spectral density using $\Delta v_i = \pi h_o$, where $h_o$ is the white frequency noise level [27], [28]. Therefore, this quantity enables a direct and meaningful comparison of the performance of different laser systems. In particular, it allows us to quantify the linewidth narrowing observed in this Raman laser system. Figures 3(a, b) show the measurement setup and the frequency noise of the pump and Stokes fields measured using a commercial frequency-noise analyzer and independently verified with a home-built correlated self-heterodyne system [28]. As shown in Fig. 3(b), the Raman lasing process in diamond reduces the pump frequency-noise spectral density from $3\times10^6$ Hz$^2$/Hz to 0.9 Hz$^2$/Hz at the highest output power of 35 W, which corresponds to a linewidth-narrowing factor of approximately $3\times10^6$ and leads to a laser intrinsic Stokes linewidth of 3 Hz and yellow linewidth of 6 Hz at 35W. Although the linewidth is measured at the Stokes wavelength, the frequency-doubled yellow light exhibits a linewidth exactly twice that of the Stokes, as dictated by the phase-matching condition, with the factor of two confirmed with high precision [29]. Such suppression exceeds by several orders of magnitude that reported for other linewidth-reduction techniques, including Brillouin lasers and injection locking [22][22], and approaches the $10^8$-level predicted by the numerical model [20]. The current measurement is limited by the noise floor of the frequency-noise analyzer, but the numerical model predicts an even narrower linewidth. Although such a narrow intrinsic linewidth is not essential for guide star applications, it is critical for quantum experiments where high-frequency noise degrades control fidelity since it matches the motional frequencies of the trapped atoms [3] [4]. The seed Gaussian linewidth is intentionally broadened to ~1.5 GHz to avoid the SBS in the pump fiber amplifier. Although the Gaussian linewidth of Stokes is governed by the technical noise of the laser cavity and cannot be used as a metric for quantifying the linewidth narrowing by Raman lasers, it is measured using delayed-self-heterodyne interferometry to be 63 kHz at the maximum power, as shown in Fig. 3 (c). The Gaussian linewidth of the yellow light is exactly twice that of the Stokes (126 kHz), as discussed above and precisely validated [29].

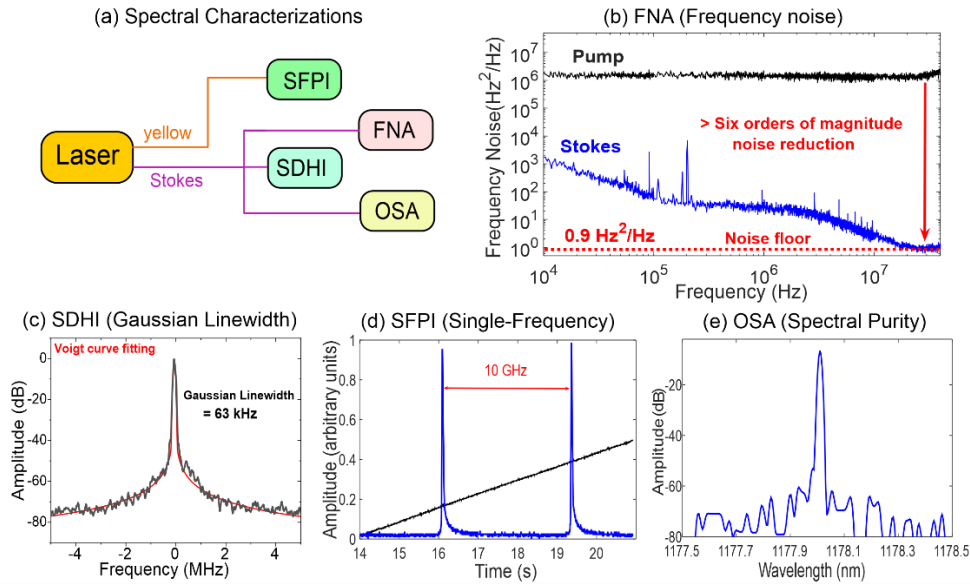

Fig. 3 Spectral characterization of the diamond Raman laser (a) Experiments used in characterization, SFPI: scanning Fabry-Perot interferometer, FNA: frequency noise analyzer, SDHI: self-heterodyne interferometer, OSA: Optical spectral analyzer laser at all power levels measured with a scanning Fabry-Perot interferometer, (b) frequency noise spectral density of the Stokes and the pump lasers, (c) Gaussian linewidth measurement using the SDHI (Self-delayed heterodyne interferometer), which reveals a Gaussian linewidth of 63 kHz (10 ms integration time), (d) an SFPI scan which demonstrates the single-mode operation of the laser, (e) spectrum trace of the laser over a 2 nm range using an Optical spectrum analyzer, which shows the laser spectral purity.

Previous reports of diamond Raman lasers indicate that mode-hop-free single-frequency operation was only possible for seconds or a maximum of 10 minutes, even at very low powers [17] [30]. Here, we measured a mode-hop-free single-mode operation at 6 W for half an hour.

The frequency stability and mode-hop-free operation of the laser are mainly limited by the thermal stability of the cavity, particularly that of the diamond crystal. Temperature variations in the diamond change both the cavity optical path length and the Raman shift. Further engineering, including low–thermal-expansion cavity materials and improved thermal management of the diamond, is therefore required to extend the mode-hop-free operating range, especially at higher powers. A scanning Fabry-Perot interferometer and an etalon-based wavemeter are used to examine the single-frequency operation of the laser at different powers. Fig. 3 (d) shows the laser mode structure at maximum power, indicating single-frequency operation. Fig. 3(e) shows an OSA trace with a 2 nm wide range, which indicates a spectrally pure laser spectrum.

## 5. Laser frequency stabilization

*5.1 Tuning range*

One of the primary challenges in stabilizing the laser was the limited mode-hop-free tuning range when sweeping a cavity mirror, which was typically less than about 400 MHz, only 25% of the cavity FSR. We attribute this limitation to parasitic reflections between each diamond face and the cavity mirrors. While wedging diamond faces can suppress these etalon effects, it significantly raises the lasing threshold and significantly reduces efficiency. To overcome this, we implemented a dual-piezo approach [31]: as shown in Fig. 4 (a), by synchronously controlling both the input coupler and end mirrors by the same gain, the two parasitic cavity modes formed between each diamond face and its corresponding mirror are coherently dithered. This prevents destructive interference between the two parasitic cavity modes, which is a potential cause of mode-hopping. This method increased the mode-hop-free tuning range to 5.5 GHz, as shown in Fig. 4(b), enabling straightforward frequency stabilization to the Na $D_{2a}$ transition needed for guide star generation.

*5.2 Sodium saturation spectroscopy*

Laser stabilization to the sodium transition is achieved via frequency modulation saturation spectroscopy [32], as shown in Fig. 4(b). The 589 nm output from the cavity end mirror is split into probe and pump beams. Most of the light is directed backwards to saturate the transition (pump), so that the forward beam sees the saturated transition at PD1. For canceling laser amplitude fluctuations and the Doppler background from the feedback signal, a second beam is sent through the cell to photodetector PD2. The differential signal between both photodiodes is fed to the laser locking electronics, which applies lock-in detection and produces the dispersive-like error signal shown in Fig. 4(c). Fig. 4 (d) shows the time-trace of the laser optical frequency at the Stokes wavelength (1178 nm). Fig. 4 (e) shows the relative Alan Standard Deviation (ADEV) of the measured Stokes frequency is $3\times10^{-9}$. The measurement is limited by the wavemeter precision for averaging times less than 10 ms, where the stability is expected to be more than one order of magnitude better based on the integrated linewidth measurement ($\Delta\nu/\nu \sim 2\times10^{-10}$; $\tau=10$ ms) in Fig. 3 (c). The laser was stabilized to the sodium D2a hyperfine transition for 30 minutes at 6 W without any mode hopping. Longer stabilization times and operation at higher powers are expected through improved thermal dissipation and cavity engineering with low thermal expansion mirror spacers.

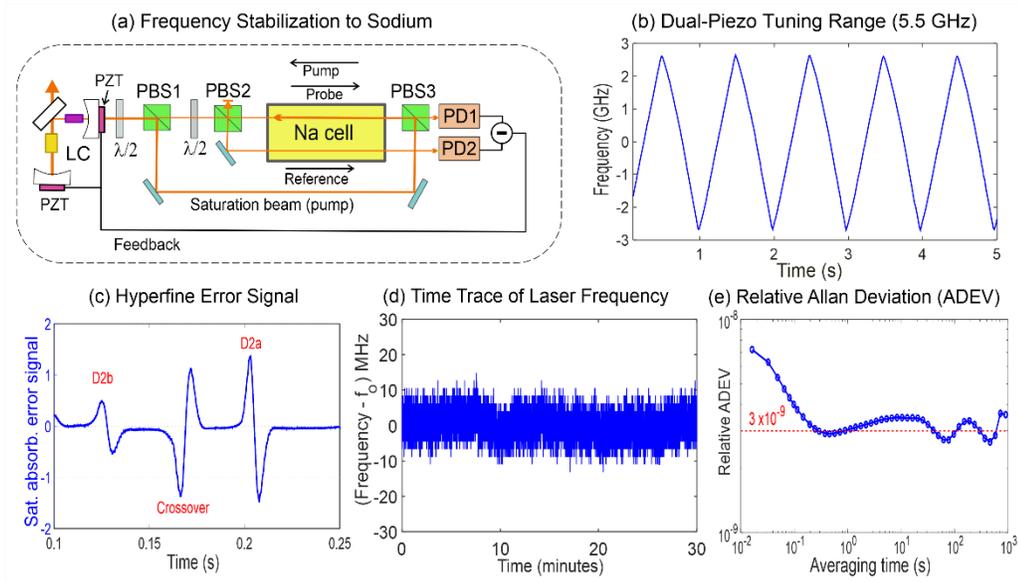

Fig. 4. Stabilization to the hyperfine D2a transition. (a) Frequency stabilization setup: a sample of the 589 nm output from the end mirror was used. The half-waveplate ($\lambda/2$) and polarization beam splitters (PBS1, PBS3) direct most of the power in the reverse direction to saturate the sodium transition. The probe light is sent to the photodetector (PD1) in the forward direction, and a reference beam to the photodetector (PD2) for Doppler background removal. (b) The 5.5 GHz laser tuning range obtained by synchronously scanning the input coupler and the end mirror, (c) locking error signal showing $D_2a$, $D_2b$, and crossover signals, which span 1.7 GHz. (d) Time trace of the optical frequency measured for the stabilized Stokes with a wavemeter at 6 W of yellow laser power ($f_0$= 254 423 970 MHz) (e) Relative Allan Deviation (ADEV) of the measured frequency stability, which shows relative stability of $3\times10^{-9}$.

## 6. Discussion and conclusion

The present work demonstrates an extraordinary linewidth narrowing factor of $3\times10^6$ enabled by Raman lasing in diamond, resulting in an intrinsic linewidth of approximately Hz-scale linewidth at an output power of 35 W. While the Stokes Gaussian linewidth is primarily determined by technical noise in the laser cavity, our measurements show that it is reduced to 63 kHz from the GHz-broadened linewidth of the pump, which is intentionally broadened to suppress SBS in the fiber amplifiers. Standard frequency-locking techniques using a reference cavity could further reduce the Gaussian linewidth, as it lies well within the control bandwidth of typical feedback electronics. In contrast, the intrinsic linewidth originates from high Fourier-frequency noise that lies beyond the bandwidth of conventional control systems and therefore cannot be easily suppressed electronically. High-frequency noise is critical for several quantum experiments, as it matches the motion frequencies of the trapped atoms, so it can degrade control fidelity and limit the sideband cooling performance [3] [4].

For space-based applications, the combination of high output power of 35 W and near-diffraction-limited beam quality ($M^2$ =1.07) is highly advantageous. Stabilizing the laser frequency to the sodium $D_2a$ hyperfine transition is also essential for reliable excitation of mesospheric sodium atoms and for maintaining stable photon return in laser guide star systems. Although extremely narrow linewidths are not required for conventional adaptive optics, recent approaches using optical chirping to mitigate spectral hole burning caused by photon recoil and saturation at high power suggest that linewidth may play an important role. Numerical modeling predicts that photon return could be enhanced by up to 60% for extremely narrow-linewidth lasers, whereas experiments with MHz-scale linewidths have demonstrated enhancements of approximately 22% [33] [34]. In addition to adaptive optics, narrow-linewidth sources at this wavelength are valuable for applications such as mesospheric sodium magnetometry and sodium-layer remote sensing, which rely on high-resolution spectroscopic interrogation of the sodium resonance. Although these applications may not require the full linewidth performance demonstrated here, the availability of such sources will enable new experiments in this field. Furthermore, the linewidth can be tuned if required, for example, through phase modulation using an electro-optic modulator. Importantly, the strong linewidth-narrowing effect inherent to Raman lasing also facilitates power scaling, as it allows the seed laser to be spectrally broadened to suppress SBS in fiber amplifiers without significantly degrading the linewidth of the final output laser.

## 7. Methods

### 7.1 Pump design

Two seed lasers were used interchangeably during the experiments. The first was a Moglabs Cateye extended-cavity laser producing ~35 mW of single-frequency output. Its linewidth was broadened using an IXBlue NIR-MX-LN-10 electro-optic phase modulator (EOM) driven by a NOISECOM UFX7112 white-noise generator, enabling SBS-free amplification in the main Yb-doped fiber amplifier (YDFA). The second seed was a more compact, portable design intended for field deployment in actual guide star laser sky tests: a distributed feedback (DFB) laser from LD-PD (Singapore) operating at 1018 nm with ~50 mW output power. Its linewidth was similarly broadened to ~1 GHz using an LD-PD EOM and a white-noise generator. Both seeds were pre-amplified to ~110 mW before injection into the main YDFA, (FiLase), which supports up to 81 W of pump power in linear polarization. The YDFA output, collimated to a 1.2 mm beam diameter, was directed into the first isolator (ISO1 in Fig. 1(a)), an N-SF1-based device from EOT rated for 400 W. The beam was then expanded twofold using a 2× beam expander (Wavelength OE) before passing through a second isolator (ISO2), a TGG-based device from CS TEC. A subsequent 3× beam expander (Wavelength OE) increased the beam diameter further, enabling tight focusing of the pump inside the diamond crystal via a 100 mm focal-length plano-convex lens.

### 7.2 Cavity design

All three mirrors are coated for high reflectivity (HR) at the Stokes field at 1178 nm. At the pump wavelength of 1018 nm, the input coupler is highly transmitting, while the end and turning mirrors are reflective for double-pass pumping. The input coupler is reflective at 589 nm, while the turning mirror is transmitting for double-pass second-harmonic generation (SHG) in the LBO crystal. The input coupler and end mirrors have a radius of curvature (ROC) of 50 mm, and the turning mirror is flat. A plano-convex lens ($f$ = 100 mm) focuses the pump beam into the diamond, producing a measured waist of 30 µm. The cavity mirrors focus the oscillating Stokes mode to a calculated waist of 45 µm at 2 mm from the stability limit. The tighter pump focusing relative to the Stokes mode promotes efficient pump depletion and reduces the lasing threshold. The LBO crystal (Crystal Laser 3x2), $4 \times 4 \times 8$ mm³, designed for type-I non-critical phase matching, is positioned less than 10 mm from the input coupler, allowing the turning mirror to be placed far enough from the focal spot to prevent coating damage at high intracavity powers. The LBO is housed in a sandwiched micro-thermoelectric cooler (TEC) for rapid, precise temperature control. A 10 kΩ thermistor on the mount is connected to a Thorlabs TED200 temperature controller, stabilizing the crystal near 40 °C. The Stokes beam waist at the LBO is ~260 µm (cavity 2 mm away from the stability limit), corresponding to the under-coupling regime described in ref [18]. Two Element Six diamond crystals, used interchangeably, measure $2 \times 2 \times 7$ mm³ and $4 \times 2 \times 7$ mm³, with similar birefringence maps (Fig. 3(d)) and nitrogen content below 100 ppb. The diamond is passively air-cooled, while the LBO is actively temperature-stabilized. The total loss in the system is calculated from the simulation to be 0.4% due to the low loss of the mirrors. The intracavity power reaches 9 kW at the maximum pump level.

*7.3 Thermal lensing mitigation*

Thermal lensing of the pump beam in the beam condition and delivery optics before the DRL was calculated and experimentally assessed by imaging the beam at the focal plane after placing all optical components in the pump path, excluding the diamond and the end mirror. Among the components, the TGG Faraday rotator in the second isolator (ISO2 in Fig. 1(b)) is expected to be the most susceptible to thermal lensing, as it experiences both the forward-propagating and back-reflected residual pump power. Calculations indicate that it could cause a focal shift of approximately –2 mm. To mitigate this, a two-fold beam expander was inserted before the second isolator. Inside the cavity, thermal lensing effects were not observed and are assumed negligible due to the relatively large beam diameter (~1.5 mm) in the LBO and the material's very low absorption, and in the case of diamond due to its exceptionally high thermal conductivity (~2000 W/m·K).

*7.4 Laser Characterization*

Several diagnostic tools were employed for laser characterization. The beam profile was measured using a Thorlabs scanning-slit beam profiler. Laser wavelength was monitored with a high-resolution wavemeter (HighFinesse WS6-200), providing an absolute accuracy of 200 MHz and repeatability of 8 MHz. Frequency noise and relative intensity noise (RIN) were measured using an OEwaves OE4000 noise analyzer, with results cross-validated via a cross-correlation self-heterodyne setup [28]. The laser linewidth was determined using a home-built self-heterodyne interferometer with a 100 MHz frequency shift, a 2 km SM980 single-mode fiber delay line, and an Anritsu optical spectrum analyzer. Single-mode operation was verified with both a Thorlabs scanning Fabry–Perot interferometer (SFPI) and the wavemeter. Diamond birefringence was measured using a home-built setup based on the method used in [26]. Output power was measured with a Thorlabs S314C thermal power sensor, capable of handling up to 2 kW/cm². For high-power chopping tests, a Thorlabs MC2000B optical chopper with a custom high-durability blade was used. A high-resolution optical spectrum analyzer (Thorlabs RedStone) enabled detection of narrow spectral features, such as SBS and FWM sidebands, located close to the main laser peak.

*7.5 Chaotic laser emission*

Chaotic laser emission is the major limitation to power scaling and arises primarily from stimulated Brillouin scattering (SBS) and four-wave mixing (FWM) in the cavity. Stimulated Brillouin scattering (SBS) originates from intracavity Stokes light and is further amplified by the Raman gain. While the fundamental SBS peaks can be avoided by slight adjustments of the cavity length, higher-order cavity spatial modes induced by SBS are more difficult to suppress [23]. In a linear cavity with a flat turning mirror, placing an aperture reduces the power of all higher-order modes. Since they are different in power along the cavity length, the aperture is effectively creating a range of cavity lengths free from SBS. In contrast, the astigmatism introduced by a spherical turning mirror causes higher-order modes to split into similar powers across the cavity, making them challenging to avoid [23]. The impact of SBS on laser performance can be characterized by modulating the pump beam with a chopper and observing Stokes pulse dynamics.

Another nonlinear effect observed is four-wave mixing (FWM), which generates weak sidebands separated from the main laser peak by multiples of the cavity free spectral range (FSR). The sideband spacing can vary unpredictably and differs from prior observations. FWM-induced chaos can be mitigated by minimizing excess intracavity Stokes power through optimized output coupling and effective suppression of SBS, as both effects are strongly correlated.

**Contributions**

O.T. conceived, planned, and led the experiments, built the laser and sodium stabilization setup, performed measurements, established the noise experiment, and wrote the manuscript. A.S. contributed to cavity design, proposed the frequency control technique, performed simulations, assisted with measurements, and contributed to writing. A.C. contributed to designing and building the pump laser. M.F. contributed to the noise measurements. D.S., J.L., and T.O. contributed through discussions. R.M. conceived the experiments and provided conceptual input, contributed to writing and reviewing, and secured funding.

**Funding**

This work is supported by the Australian Research Council (Grant No. LP200301594), the U.S. Air Force Office of Scientific Research (Grant No. FA2386-21-1-4030), and EOS Space Pty Ltd.

**Data availability.** The data underlying the results presented in this paper are available on request.

**Disclosure:** The authors declare no conflicts of interest.

**References**

[1] K. Xu *et al.*, 'Sodium Bose-Einstein condensates in an optical lattice', *Phys. Rev. A*, vol. 72, no. 4, Oct. 2005, doi: 10.1103/PhysRevA.72.043604.

[2] W.-Y. Zhang *et al.*, 'Scalable Multipartite Entanglement Created by Spin Exchange in an Optical Lattice', *Phys. Rev. Lett.*, vol. 131, no. 7, p. 73401, Aug. 2023, doi: 10.1103/PhysRevLett.131.073401.

[3] L. Krinner, K. Dietze, L. Pelzer, N. Spethmann, and P. O. Schmidt, 'Low phase noise cavity transmission self-injection locked diode laser system for atomic physics experiments', *Opt. Express*, vol. 32, no. 9, p. 15912, Apr. 2024, doi: 10.1364/oe.514247.

[4] M. L. Day, P. J. Low, B. White, R. Islam, and C. Senko, 'Limits on atomic qubit control from laser noise', *npj Quantum Inf.*, vol. 8, no. 1, Dec. 2022, doi: 10.1038/s41534-022-00586-4.

[5] C. E. Max *et al.*, 'Image Improvement from a Sodium-Layer Laser Guide Star Adaptive Optics System', *Science (1979).*, vol. 277, no. 5332, pp. 1649–1652, Sep. 1997, doi: 10.1126/science.277.5332.1649.

[6] C. D'Orgeville *et al.*, 'A sodium laser guide star facility for the ANU/EOS space debris tracking adaptive optics demonstrator', in *Adaptive Optics Systems IV*, SPIE, Jul. 2014, p. 91483E. doi: 10.1117/12.2055050.

[7] R. Mata-Calvo *et al.*, 'Laser guide stars for optical free-space communications', in *Free-Space Laser Communication and Atmospheric Propagation XXIX*, SPIE, Feb. 2017, p. 100960R. doi: 10.1117/12.2256666.

[8] T. Li, X. Fang, W. Liu, S.-Y. Gu, and X. Dou, 'Narrowband sodium lidar for the measurements of mesopause region temperature and wind', *Appl. Opt.*, vol. 51, no. 22, pp. 5401–5411, Aug. 2012, doi: 10.1364/AO.51.005401.

[9] J. M. Higbie, S. M. Rochester, B. Patton, R. Holzlöhner, D. B. Calia, and D. Budker, 'Magnetometry with mesospheric sodium', *Proc. Natl. Acad. Sci. U. S. A.*, vol. 108, no. 9, pp. 3522–3525, Mar. 2011, doi: 10.1073/pnas.1013641108.

[10] X. Huo *et al.*, 'Research development of 589 nm laser for sodium laser guide stars', *Opt. Lasers Eng.*, vol. 134, p. 106207, 2020, doi: https://doi.org/10.1016/j.optlaseng.2020.106207.

[11]  D. B. Calia, Y. Feng, W. Hackenberg, R. Holzlöhner, L. Taylor, and S. Lewis, 'Telescopes and Instrumentation Laser Development for Sodium Laser Guide Stars at ESO', 2010.

[12]  Y. Feng, L. R. Taylor, and D. B. Calia, '25 W Raman-fiber-amplifier-based 589 nm laser for laser guide star', *Opt. Express*, vol. 17, no. 21, pp. 19021–19026, 2009, doi: 10.1364/OE.17.019021.

[13]  D. B. Calia *et al.*, 'Novel 63W CW 589nm chirped laser for laser guide star adaptive optics', in *Adaptive Optics Systems VIII*, L. Schreiber, D. Schmidt, and E. Vernet, Eds., SPIE, 2022, p. 121852X. doi: 10.1117/12.2630398.

[14]  M. Zhang *et al.*, 'Frequency-Stabilized High-Power 589 nm Semiconductor Disk Laser for Guide Star Applications', in *Laser Congress 2024 (ASSL, LAC, LS&C)*, Optica Publishing Group, 2024, p. ATu1A.4. doi: 10.1364/ASSL.2024.ATu1A.4.

[15]  G. J. Fetzer, J. Chilla, S. E. Rako, C. Baumgarten, N. Woody, and C. D'Orgeville, 'High-power single frequency intracavity doubled VECSEL at 589 nm for sodium guide stars', SPIE-Intl Soc Optical Eng, Mar. 2022, p. 16. doi: 10.1117/12.2612400.

[16]  K. S. Gardner, R. H. Abram, and E. Riis, 'A birefringent etalon as single-mode selector in a laser cavity', *Opt. Express*, vol. 12, no. 11, pp. 2365–2370, May 2004, doi: 10.1364/OPEX.12.002365.

[17]  A. Sharp *et al.*, 'Isolator-free 60 W diamond Raman laser at 607 nm', *Opt. Lett.*, vol. 49, no. 18, p. 5139, Sep. 2024, doi: 10.1364/ol.538377.

[18]  H. Jasbeer, R. J. Williams, O. Kitzler, A. McKay, and R. P. Mildren, 'Wavelength diversification of high-power external cavity diamond Raman lasers using intracavity harmonic generation', *Opt. Express*, vol. 26, no. 2, p. 1930, Jan. 2018, doi: 10.1364/oe.26.001930.

[19]  X. Yang, O. Kitzler, D. J. Spence, Z. Bai, Y. Feng, and R. P. Mildren, 'Diamond sodium guide star laser', *Opt. Lett.*, vol. 45, no. 7, p. 1898, Apr. 2020, doi: 10.1364/ol.387879.

[20]  R. L. Pahlavani, D. J. Spence, A. O. Sharp, and R. P. Mildren, 'Linewidth narrowing in Raman lasers', *APL Photonics*, vol. 10, no. 7, Jul. 2025, doi: 10.1063/5.0271652.

[21]  E. Granados *et al.*, 'Spectral synthesis of multimode lasers to the Fourier limit in integrated Fabry–Perot diamond resonators', *Optica*, vol. 9, no. 3, p. 317, Mar. 2022, doi: 10.1364/optica.447380.

[22]  M. Deroh, E. Lucas, K. Hammani, G. Millot, and B. Kibler, 'Linewidth-narrowing and frequency noise reduction of Brillouin fiber laser cavity operating at 1-µm', *EPJ Web Conf.*, vol. 287, p. 09003, 2023, doi: 10.1051/epjconf/202328709003.

[23]  M. Li, D. J. Spence, Y. Sun, X. Yang, and Y. Feng, 'Spatially controllable stimulate Brillouin scattering in the diamond Raman laser', *Opt. Laser Technol.*, vol. 181, Feb. 2025, doi: 10.1016/j.optlastec.2024.111979.

[24] A. Sharp, O. Terra, D. J. Spence, O. Kitzler, R. Pahlavani, and R. P. Mildren, 'Phase offset and polarisation flipping in diamond Raman lasers with intracavity second harmonic generation', *Opt. Express*, vol. 33, no. 23, pp. 48068–48080, Nov. 2025, doi: 10.1364/OE.575557.

[25] S. Antipov *et al.*, 'Analysis of a thermal lens in a diamond Raman laser operating at 1.1 kW output power', *Opt. Express*, vol. 28, no. 10, pp. 15232–15239, May 2020, doi: 10.1364/OE.388794.

[26] H. Jasbeer *et al.*, 'Birefringence and piezo-Raman analysis of single crystal CVD diamond and effects on Raman laser performance', *Journal of the Optical Society of America B*, vol. 33, no. 3, p. B56, Mar. 2016, doi: 10.1364/josab.33.000b56.

[27] G. Di Domenico, S. Schilt, and P. Thomann, 'Simple approach to the relation between laser frequency noise and laser line shape', 2010.

[28] Z. Yuan *et al.*, 'Correlated self-heterodyne method for ultra-low-noise laser linewidth measurements', *Opt. Express*, vol. 30, no. 14, p. 25147, Jul. 2022, doi: 10.1364/oe.458109.

[29] J. Stenger, H. Schnatz, C. Tamm, and H. R. Telle, 'Ultraprecise Measurement of Optical Frequency Ratios', *Phys. Rev. Lett.*, vol. 88, no. 7, p. 4, 2002, doi: 10.1103/PhysRevLett.88.073601.

[30] Y. Liu *et al.*, 'High-power free-running single-longitudinal-mode diamond Raman laser enabled by suppressing parasitic stimulated Brillouin scattering', *High Power Laser Science and Engineering*, vol. 11, Aug. 2023, doi: 10.1017/hpl.2023.67.

[31] A. Fawzy, O. M. El-Ghandour, and H. F. A. Hamed, 'Performance Analysis on a Dual External Cavity Tunable Laser ECTL Source', *Journal of Electromagnetic Analysis and Applications*, vol. 07, no. 04, pp. 134–139, 2015, doi: 10.4236/jemaa.2015.74015.

[32] J. Ding *et al.*, 'Compact and high reliable frequency-stabilized laser system at 589 nm based on the distributed-feedback laser diodes', *Appl. Phys. B*, vol. 127, no. 9, Sep. 2021, doi: 10.1007/s00340-021-07677-8.

[33] F. Pedreros Bustos, R. Holzlöhner, S. Rochester, D. Bonaccini Calia, J. Hellemeier, and D. Budker, 'Frequency chirped continuous-wave sodium laser guide stars: modeling and optimization', *Journal of the Optical Society of America B*, vol. 37, no. 4, p. 1208, Apr. 2020, doi: 10.1364/josab.389007.

[34] J. Hellemeier *et al.*, 'Laser guide star return-flux gain from frequency chirping', *Mon. Not. R. Astron. Soc.*, vol. 511, no. 3, pp. 4660–4668, Apr. 2022, doi: 10.1093/mnras/stac343.